\documentclass[prd,superscriptaddress,preprintnumbers,nofootinbib,showkeys,notitlepage,twocolumn]{revtex4-1}
\usepackage{graphicx,amssymb,amsmath,color}
\usepackage{hyperref}
\usepackage{multirow}
\usepackage{ulem}

\newcommand{\beq}{\begin{equation}}
\newcommand{\eeq}{\end{equation}}
\def\bea{\begin{eqnarray}}
\def\eea{\end{eqnarray}}
\newcommand{\bei}{\begin{itemize}}
\newcommand{\eei}{\end{itemize}}

\newcommand{\Fig}[1]{Fig.~\ref{#1}}
\newcommand{\Eq}[1]{Eq.~(\ref{#1})}

\newcommand{\App}[1]{Appendix~\ref{#1}}

\def\={\,=\,}
\def\+{\,+\,}
\def\-{\,-\,}
\def\Msun{M_{\odot}}
\def\rsun{r_{\odot}}

\begin{document}

\title{
GRB lensing parallax: Closing primordial black hole dark matter mass window
}

\author{Sunghoon Jung}
\email{sunghoonj@snu.ac.kr}
\affiliation{Center for Theoretical Physics, Department of Physics and Astronomy, Seoul National University, Seoul 08826, Korea}

\author{TaeHun Kim}
\email{gimthcha@snu.ac.kr}
\affiliation{Center for Theoretical Physics, Department of Physics and Astronomy, Seoul National University, Seoul 08826, Korea}


\begin{abstract}

The primordial black hole (PBH) comprising full dark matter (DM) abundance is currently allowed if its mass lies between $10^{-16}\Msun \lesssim M \lesssim 10^{-11} \Msun$. This lightest mass range is hard to be probed by ongoing gravitational lensing observations. In this paper, we advocate that an old idea of the lensing parallax of Gamma-ray bursts (GRBs), observed simultaneously by spatially separated detectors, can probe the unconstrained mass range; and that of nearby stars can probe a heavier mass range. In addition to various good properties of GRBs, astrophysical separations achievable around us -- $r_\oplus \sim$ AU -- is just large enough to resolve the GRB lensing by lightest PBH DM. 

\end{abstract}


\maketitle


\section{Introduction}
The PBH has become an interesting DM candidate. Possibly having been formed from overdensities in primordial density fluctuations during inflation, it might have evaporated too much by today if it is too light or could have disturbed CMB if too heavy. A wide mass range between those general constraints remains open as DM.

PBH DM have been probed mainly by two categories: gravitational lensing and PBH-induced dynamics. The latter includes dynamical disruption, capture, and friction of observed systems, but are subject to various assumptions and uncertainties~\cite{Carr:2016drx,Sasaki:2018dmp}. Maybe the lensing is a more robust and direct probe. Among several proposals, basically only microlensing of nearby stars have been thoroughly measured, constraining $10^{-11}\Msun \sim 10 \, \Msun$~\cite{Niikura:2017zjd,Niikura:2019kqi,Griest:2013aaa,Tisserand:2006zx,Allsman:2000kg}; see also~\cite{Dror:2019twh}. Heavier PBHs are expected to be probed by strong lensing of supernova~\cite{Zumalacarregui:2017qqd}/FRB~\cite{Munoz:2016tmg}/GRB and lensing fringes of GW at aLIGO~\cite{Jung:2017flg}. On the other hand, lensing cannot efficiently probe lighter PBHs in the range $10^{-16} \Msun$ -- $10^{-11} \Msun$, as lensing is typically too weak and fragile to be resolved; in particular, source size and wave-optics effects are limitations~\cite{Katz:2018zrn,Bai:2018bej}. 

In this paper, we study and advocate lensing parallaxes of GRBs and stars as important probes of PBH DM, with a potential to probe the whole possible mass range with lensing.

\section{Lensing Parallax} 
Lensing parallax allows the detection of lensing by correlating the brightnesses (or fluxes) of a source simultaneously measured by spatially separated detectors~\cite{Nemiroff:1995ak}. If the spatial separation is larger than about the Einstein radius of a PBH lens, the two detectors will observe different lensing magnifications. The observed magnification of unresolved images is
\beq
A \= \mu_1 \+ \mu_2 \+ \sqrt{\mu_1 \mu_2} \,\cos 2\pi c\Delta t_d/\lambda,
\label{eq:magnification}  \eeq
with each image's magnification
\beq
\mu_i \= \left| \tfrac{1}{2} \,\pm \, (y^2 +2)/[2y\sqrt{y^2+4}] \right|,
\label{eq:mageach} \eeq
and time-dealy between images $\Delta t_d$.
\Eq{eq:mageach} is applicable for a point-lens and source in the geometrical optics, but will be corrected by source-size effects in final results; see below \Eq{eq:epsilon} and \App{app:sourcesize}. 
It is this relative angle $y$ of a source with respect to a lens (normalized by the lens Einstein angle) that can differ between spatially separated detectors by order one, causing lensing parallax. Since the interference fringe term quickly averages out, magnification becomes $A \simeq \mu_1+\mu_2$, frequency-independent for the majority of parameter space. Thus, different brightness with the same energy spectra between simultaneous measurements is a basic signal of lensing. However, the interference fringe (in the energy spectrum)~\cite{Gould:1992,Nakamura:1997sw,Jung:2017flg} becomes relevant when the wavelength $\lambda$ is about to be too long to probe small PBHs with mass $M$, i.e. $\lambda \sim {\cal O}(GM/c^2)$ (equivalently, $c\Delta t_d/\lambda \sim {\cal O}(1)$). In this regime, frequency-dependent magnification (i.e., fringe and wave-optics effects) can further strengthen lensing-parallax detection. Our study is mainly based on the former feature (frequency-independent), but the latter is also relevant near the lower mass limit.

Lensing parallax is similar to microlensing, in that brightness is measured at multiple relative angles to detect lensing. But lensing parallax can be detected even with short transients, based on {\it simultaneous} observations from well-separated detectors (whereas microlensing needs long-lasting sources). This allows us to use GRB pulse in lensing parallax~\cite{Nemiroff:1995ak}; as will be discussed, GRB is ideal to probe {\it lightest} unconstrained PBHs. Using pulses, we assume lensing effects do not change during measurements. The correlation can be made unambiguously between a whole time-series of pulse data, but longer temporal variation can also be cross-correlated to further strengthen the detection. The lensing parallax of long-lasting stars can also be detected in either way. We assume to use only short time-segment in our estimations.

Lensing parallax requires the following conditions to be met. First, the detector separation, $\Delta r$, should be larger than about the Einstein radius, $r_E$, of a PBH lens:
\beq
\Delta r \, \gtrsim \, r_E   \quad  \Leftrightarrow \quad  \left(\frac{M}{10^{-7} \Msun} \right) \, \lesssim \, \left( \frac{\Delta r}{\rm AU} \right)^2 \left( \frac{D}{\rm Gpc} \right)^{-1},
\label{eq:max} \eeq
where typical distance to the lens and source is denoted by $D \sim D_S \sim D_L$. A proportionality constant is not shown but varies orders of magnitudes depending on the fractional brightness resolution $\epsilon$, determined by \Eq{eq:epsilon}. This condition determines the maximum PBH mass that can be probed -- $M_{\rm max}$.

Second, the apparent source angular size, $\theta_S \sim r_S/D$ (where $r_S$ is the physical transverse emission size), must be smaller than about the Einstein angle $\theta_E$ to induce sizable lensing; otherwise, only a small part of the source will be magnified. 
Requiring measured magnifications $A_i$ to be more different than the resolution,
\beq
\delta A \,\equiv\, \frac{|A_1 - A_2|}{(A_1+A_2)/2} \, \gtrsim \, \epsilon,
\label{eq:epsilon} \eeq
where each $A_i$ is given by \Eq{eq:magnification}, and we assume to compare only two measurements for simplicity. We take into account finite source size following Ref.~\cite{Witt:1994} (see also \App{app:sourcesize}), where constant brightness profile and geometrical optics are used. The latter assumption is fine in our analysis as will be discussed; and the former, compared to profiles having a peak in the center, may slightly overestimate or underestimate the magnification depending on $y$.
Source size $r_S$ taken into account, the maximum magnification is $A_{\rm max} \=  \sqrt{1 + 4/\delta^2 } \, \sim \, \delta A$ (for $y=0$ in the geometrical optics) with $\delta \, \equiv \, \theta_S / \theta_E \, \sim \, r_S/r_E$.
Thus, we obtain (in the limit $\delta \gtrsim 1$)
\beq
\left( \frac{M}{10^{-12} \Msun} \right) \, \gtrsim \, \epsilon \, \left( \frac{D}{\rm Gpc} \right)^{-1} \left( \frac{r_S}{r_\odot} \right)^2.
\label{eq:minsource} \eeq
\Eq{eq:epsilon} is the main condition that we solve numerically. 

Last but not least, the probe wavelength $\lambda$ must be shorter than the Schwarzschild radius of a PBH lens; otherwise, the wave cannot see the lens. The geometrical optics is valid for
\beq
0.1\,\lambda \, \lesssim \, \frac{2GM}{c^2}(1+z_L) \, \simeq \, 10^{-5} \, {\rm sec}\, \left( \frac{M(1+z_L)}{\Msun} \right),
\label{eq:waveoptics} \eeq
where the approximate factor 0.1 is from Ref.~\cite{Takahashi:2003ix} and $z_L$ is the lens redshift. For GRBs with $f=10^{19}-10^{21}$ Hz (50 keV -- 5 MeV) and negligible $z_L$, this condition reads $M \, \gtrsim \, (10^{-15} - 10^{-17}) \, \Msun$; for FRBs with $10^8 - 10^{10}$ Hz, $M \, \gtrsim \, (10^{-4} - 10^{-6}) \, \Msun$; and for IR-optical observations of nearby stars with $10^{14} - 10^{16}$ Hz,  $M \, \gtrsim \, (10^{-10} - 10^{-12}) \, \Msun$. We will discuss later how we treat wave optics effects.

The stronger of the last two conditions (apparent source size and wavelength) determines the minimum PBH mass that can be probed -- $M_{\rm min}$. The two constraints are similar for GRBs, the latter is stronger for FRBs (as radio wavelength is much longer than gamma-ray's), and the former is somewhat stronger for stars (as they are closer, they appear larger, and also sources are assumed to be larger).

\section{GRB Parameters}
We turn to estimate the sensitivities of GRB lensing parallax. We describe source parameters, how we treat wave-optics effects, and derive upper limits on the PBH abundance; see \App{app:GRBparameter} for more details.

The physical transverse size $r_S$ of emission region of GRB sources is typically correlated with the measured minimum variability time scale $t_{\rm var}$ as $r_S \sim c\, t_{\rm var} \cdot \Gamma/(1+z_S) \sim c\, T_{\rm 90}/(1+z_S)$~\cite{Barnacka:2014yja}, with the observed burst duration $T_{\rm 90}$, Lorentz boost $\Gamma$ and source redshift $z_S$. Based on GRBOX database of $\sim 2000$ GRBs detected so far~\cite{GRBdist}, 10\% of GRBs have $r_S \lesssim r_\odot$; 3\% have smaller $r_S \lesssim 0.1 \, r_\odot$; and the remaining 90\% have $r_S \gtrsim r_\odot$. Although constituting a minor fraction, these smallest GRBs with $r_S \lesssim r_\odot$ are the ones that allow to probe lightest unconstrained PBH mass range. It is also notable that actual GRB sizes could be smaller; a recent work estimated smaller $t_{\rm var}$ based on broader GRB spectrum~\cite{Golkhou:2015lsa}, low sampling frequency and limited photon statistics might have overestimated $t_{\rm var}$~\cite{Barnacka:2014yja}, theoretical minimum from emission region being optically thin can also be small~\cite{Katz:2018zrn}. As a concrete example, we use the GRBOX distribution.

The GRB frequency spectrum is also relevant, as the highest frequency can probe the lightest PBHs. Rather than modeling the frequency spectrum of photon counts (which varies among GRBs) and carrying out frequency-dependent lensing analysis (needed only when \Eq{eq:waveoptics} is not satisfied), we can simply consider a highest-frequency range where assumed sensitivity $\epsilon=0.1$ (conservative scenario) or 0.01 (optimistic) can be typically achieved. This is a good simplification to estimate detection prospects; if the same magnification is shown at a lower-energy range (with higher flux) or if frequency-dependent magnification is observed consistent with fringe, the lensing (fringe) detection will be strengthened. But if this highest frequency range does not show evidence of lensing, detectable lensing did not occur.

Given the typical spectrum measured by Fermi GBM+LAT~\cite{TypicalSpectrum}, we assume to use $\sim 0.5 $ MeV as the highest frequency with $\epsilon=0.1$ (conservative); this is the highest frequency where ${\cal O}(100)$ photon counts per MeV is typically achieved. This frequency range allows us to use geometrical optics down to $M_{\rm DM} \gtrsim 10^{-16} \Msun$, according to \Eq{eq:waveoptics}. In the final results, we show results only down to this mass range and defer a dedicated wave-optics study below this range. But our geometrical-optics calculation can already cover a whole unconstrained mass range. For an optimistic scenario, we assume $\epsilon=0.01$ up to $\sim 5$ MeV (hence, $M_{\rm DM} \gtrsim 10^{-17} \Msun$). Although we use geometrical optics everywhere in this simplified analysis, we take into account source-size effects as discussed. 

GRBs are at cosmological distances, observed up to $z_S \leq 10$ and majority at $z_S = 0.5 \sim 3$. We use the redshift distribution from the GRBOX database~\cite{GRBdist} to calculate lensing probability. But we do not correlate distances with observed intensities; rather, for all GRBs, we will use common intensity resolution $\epsilon$ assumed above.

Benchmark detector separations are $\Delta r = 2 r_\oplus \simeq 0.04$ sec, $r_{L2} \simeq 5$ sec, and 2 AU $\simeq 10^3$ sec. The first choice is based on telescopes on the ground and low Earth orbits such as Fermi and Swift; the second may be possible as IR-band space missions such as Gaia, WFIRST and Euclid are operating or planned on orbits around the Lagrange point $L_2$ of the Sun-Earth system; and the AU is a characteristic scale of the solar system, as is realized in Kepler mission. The detection benchmarks are summarized in Table.~\ref{tab:benchmark}.

\section{GRB Results} 
\begin{figure}[t] \centering
\includegraphics[width=0.48\textwidth]{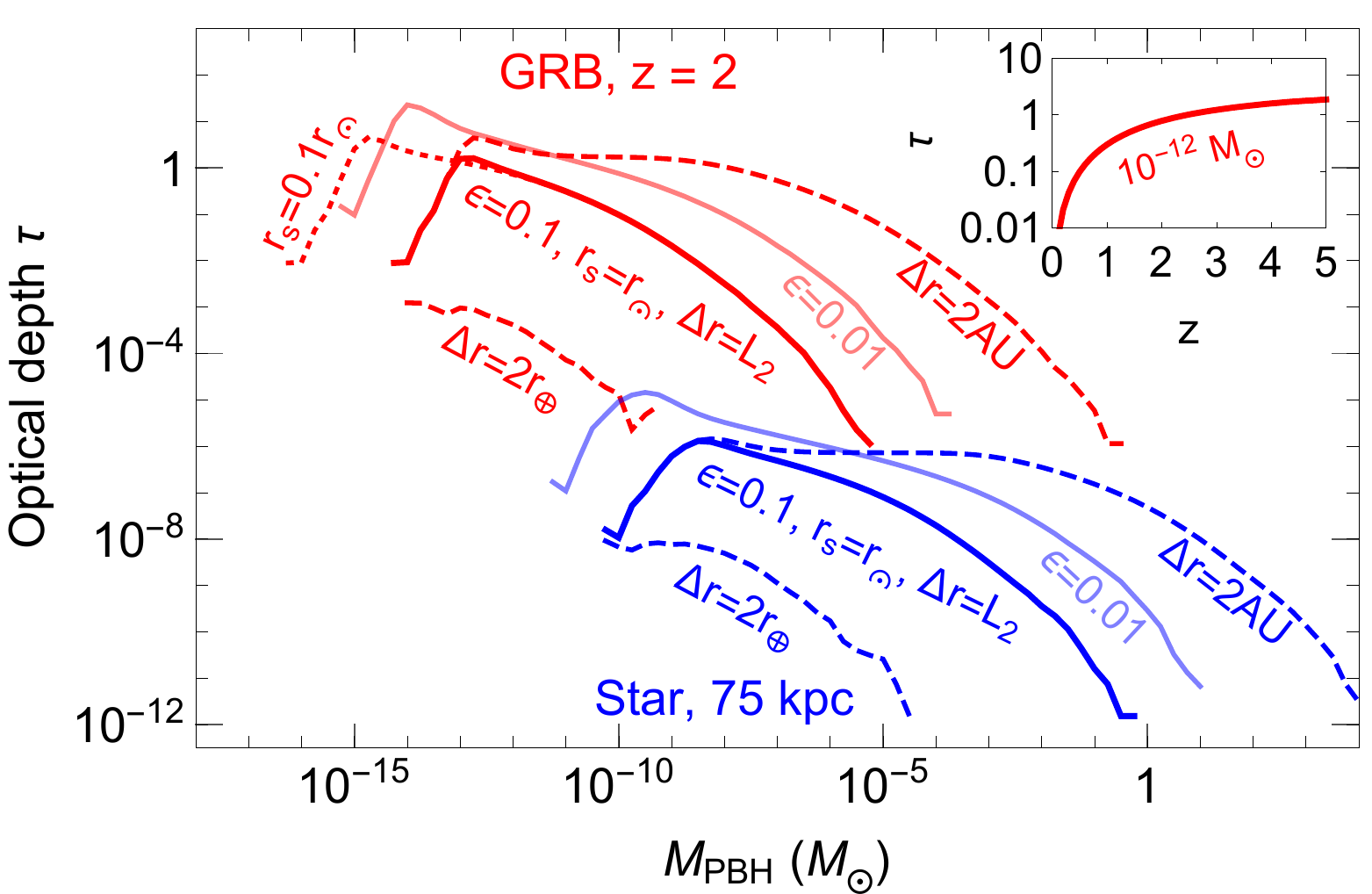}
\caption{Optical depth $\tau$ for the lensing parallax of GRBs(red) at $z_S=2$ and stars(blue) at 75 kpc. The inset shows the $z_S$-dependence for $M=10^{-12}\Msun$ as a reference. Uniform PBH distribution is used.} 
\label{fig:tau}
\end{figure}

The optical depth, $\tau$, is obtained from the volume of the desired PBH locations (for the given GRB) that can lead to appreciable lensing parallax; then the expected number of PBHs within this volume is the optical depth. PBHs are assumed to be uniformly distributed with the comoving number density $n=\rho_{crit, 0} \Omega_{\rm DM} f / M$, where $f \equiv \Omega_{\rm PBH} / \Omega_{\rm DM}$ is the PBH abundance; the clustering of PBH does not matter as long as their Einstein angles do not overlap, which is true for most regions of galaxies, as discussed in \App{app:clustering}. As shown in \Fig{fig:tau}, $\tau$ is usually smaller than 1, but $\tau \gtrsim 1$ is sizable for GRBs at far distances and for small $\epsilon$. In this case, multiple lensing can occur. Although multi-lensing can still lead to lensing parallax, we (conservatively) focus on single-lensing cases which are much easier to calculate. The single-lensing probability is $P_1 = \tau e^{-\tau}$ for the Poisson distribution of lenses, and it becomes its maximum value $0.37$ at $\tau=1$, which is still sizable.

\begin{table}[t] \centering
\begin{tabular}{ c || c  c   c c }
\hline \hline
                     &  $\epsilon$  & highest frequency & $N_{\rm GRB}$ & $\Delta r$ \\
\hline \hline 
conservative &  0.1 & $\sim$ 0.5 MeV  &  $10^3$  &  $2 r_\oplus, \, r_{L_2}, \, $ 2AU\\
\hline
optimistic      &  0.01 & $\sim$ 5 MeV &  $10^4$ &  $2 r_\oplus, \, r_{L_2}, \, $ 2AU\\
\hline \hline
\end{tabular}
\caption{Benchmarks for GRB lensing-parallax detection. Fractional brightness resolution $\epsilon$, the highest frequency used in our analysis, number of GRB detections $N_{\rm GRB}$, and detector separation $\Delta r$. For stars, $N_{\rm star}=10^7$ and $10^9$.}
\label{tab:benchmark} \end{table}

\begin{figure}[t] \centering
\includegraphics[width=0.48\textwidth]{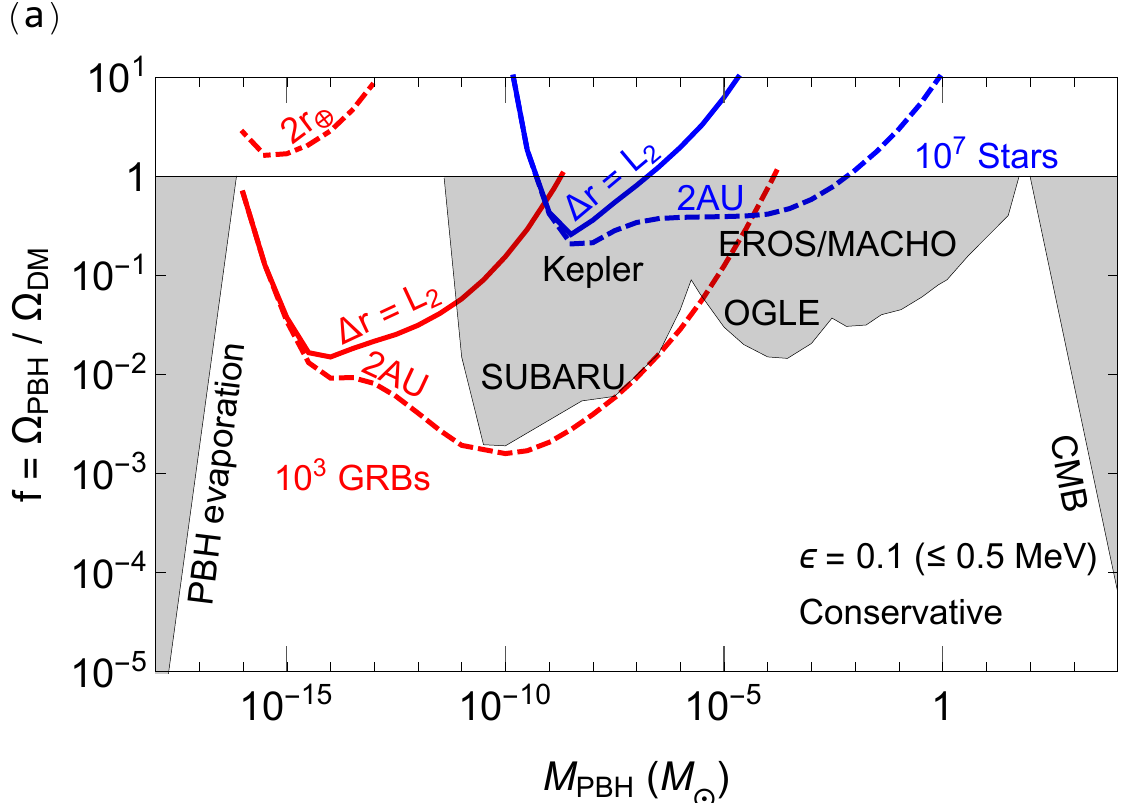}
\includegraphics[width=0.48\textwidth]{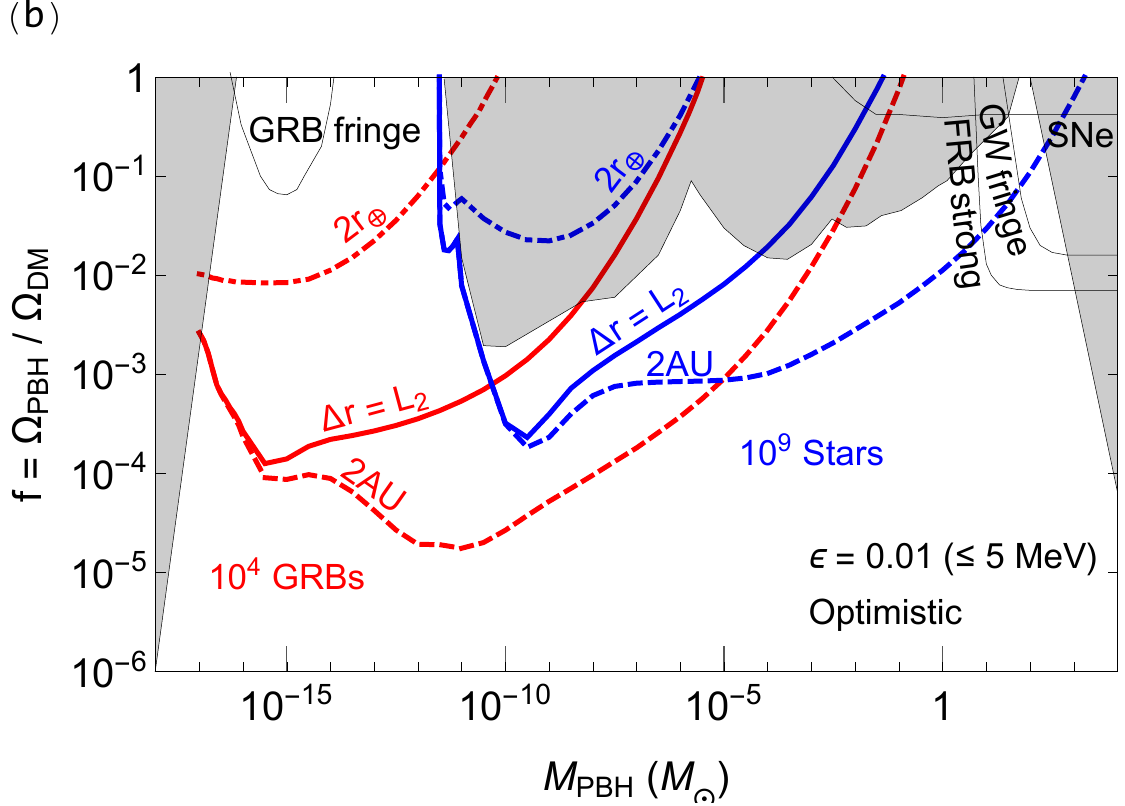}
\caption{Expected 95\% CL upper limits on the PBH abundance, $f$, from lensing parallaxes of GRBs (red) and stars (blue). The (a) upper and (b) lower panels show conservative and optimistic projections with detection benchmarks in Table~\ref{tab:benchmark}.
The region with $f>1$ is not possible but is shown for reference.
Also shown are existing (shaded) and proposed (non-shaded) lensing constraints from microlensing (SUBARU~\cite{Niikura:2017zjd}, Kepler~\cite{Griest:2013aaa}, OGLE/EROS/MACHO~\cite{Niikura:2019kqi, Tisserand:2006zx, Allsman:2000kg}), strong lensing of FRB~\cite{Munoz:2016tmg}/supernova~\cite{Zumalacarregui:2017qqd}, GW fringe (aLIGO)~\cite{Jung:2017flg}, and GRB fringe~\cite{Katz:2018zrn}; evaporation of too light PBHs~\cite{Carr:2009jm}, and CMB distortion from too heavy ones~\cite{Ali-Haimoud:2016mbv}. In all, the full range of possible PBH DM mass can be probed by lensing.} 
\label{fig:results}
\end{figure}

In \Fig{fig:results}, we obtain 95\% CL upper limits on the PBH abundance by assuming that no lensing parallax will be detected among total $N_{\rm GRB} = 1000$ (as already more than this) or $10^4$ GRBs simultaneously observed by $\Delta r$-separated detectors. The Poisson distribution of the number of lenses is used. In each of conservative and optimistic projections, three benchmark results with $\Delta r$ are shown. As discussed, we plot results down to the lowest mass where geometrical optics is valid for the highest-frequency range that we consider. 

One can first understand overall behaviors of the results. The three $\Delta r$ lines have similar $M_{\rm min}$ as it is determined dominantly by $\epsilon$ and $r_S$; $M_{\rm min} \sim 10^{-15}- 10^{-16} \Msun$ is indeed expected from \Eq{eq:minsource} with $\epsilon=0.1$, $r_S \lesssim 0.1 \, r_\odot$, and $D \sim 1$ Gpc. Optimistic results improve $M_{\rm min} \propto \epsilon$ with higher frequencies available. But the three lines have different $M_{\rm max} \propto (\Delta r)^2$ (\Eq{eq:max}). The constraints on $f$ is also improved roughly with the increase of $N_{\rm GRB}/\epsilon$, yielding higher statistics.


As a result, the unconstrained PBH DM mass range -- $10^{-16}\Msun \lesssim M \lesssim 10^{-11}\Msun$ -- can be potentially probed by lensing. Existing telescopes (having conservative specs while running on the ground and low Earth orbits) alone can barely achieve sensitivities yet; see $\Delta r=2 r_\oplus$ result. But, remarkably, one space telescope at a larger orbit ($L_2$ or AU), paired with existing ones, can be enough to probe a whole unconstrained mass range; see $\Delta r= r_{L_2}$ and 2AU results. With optimistic benchmarks, any combinations of well-separated detectors (even low-orbit ones) can probe the whole unconstrained mass range.

Confidently probing the lightest mass range near $10^{-16}\Msun$ likely requires improvements from the conservative scenario; in particular, better resolution $\epsilon < 0.1$. Such can perhaps be achieved with higher photon counting rates from larger detector area, higher sampling frequency, and better detection efficiency. GRB source size is also a key factor (\Eq{eq:minsource}) as recently emphasized for GRB fringes~\cite{Katz:2018zrn} (their result with $10^4 \times 3\% = 300$ GRBs with $r_S=10^9\,{\rm cm}$~\cite{Katz:2018zrn} is shown in \Fig{fig:results}). If sources turn out to be larger than the estimate used here, much better $\epsilon$ is needed.

Toward the heaviest mass range, GRB lensing parallax can probe potentially up to $M\lesssim 0.1 \Msun$. In the heavier mass range, various strong and microlensing observables are available, as shown in \Fig{fig:results}. Supernova lensing magnification can be detected as a deviation in the brightness-redshift relation (result with current data is shown~\cite{Zumalacarregui:2017qqd}); FRB and GRB pulses can be strongly time-delayed or angular separated ($10^4$ FRBs up to $z\leq 0.5$~\cite{Munoz:2016tmg}); and GW lensing fringes in chirping waveforms can be detected at aLIGO ($\sim 10^3$ mergers for 1 yr~\cite{Jung:2017flg}) probing the mass range corresponding to the LIGO frequency band.

\subsection{Nearby-Stars Results}
We also estimate the lensing parallax of stars in the Milky-Way and nearby galaxies. Although it cannot probe the unconstrained mass range, star's lensing parallax can perhaps be observed with ongoing and near-future missions.

We simply assume that all stars have $r_S = r_\odot$ and we use geometrical optics. We consider $10^7$ (as in current microlensings) or $10^9$ stars observed within a 100 kpc radius around us. Such a large number of sources can compensate the small optical depth ($\tau \lesssim 10^{-5}$ shown in \Fig{fig:tau}, similar to microlensing values~\cite{Niikura:2017zjd}). DM comprises the total galaxy mass $10^{12} \Msun$ and are isotropically distributed around us. 

Although stars are long-lasting, we assume to consider short time-segments as discussed; but annual parallaxes in  microlensing observations have been studied, e.g.~\cite{Refsdal:1986,Bachelet:2018}. We simply require the same lensing-detection criteria and detector benchmarks as in the GRB case; but prospects for realizing various $\Delta r$ and good resolution $\epsilon$ are higher here, as various IR-band space missions are already operating with such specs or planned to be achieving them.

The sensitivities are overlapped in \Fig{fig:results}. Stars are sensitive to heavier PBHs; as they are closer than GRBs, they appear larger and the Einstein radius of a PBH is smaller. Compared to GRBs, $M_{\rm max}$ is larger by roughly (1 Gpc/100 kpc) $\sim 10^4$ (see \Eq{eq:max}) while $M_{\rm min}$ is larger by $\sim 10^5 - 10^6$ as no stars are assumed to be smaller than the solar size. Notably, ongoing and proposed missions on the ground and in large orbits ($L_2$ or AU) can already achieve any of conservative results, probing $10^{-9}\Msun \lesssim M \lesssim 10^{-2} \, \Msun$. Optimistically, $10^{-12}\Msun \lesssim M \lesssim 10^3 \Msun$ can all be probed, complementary to microlensing and strong lensing.

\section{Discussion}
The GRB has unique advantages in completing the lensing probe of PBH DM. The large distance and compact source make it appear small compared to the Einstein angle of the lightest PBH. High-frequency gamma rays not only probe small PBHs but make the lensing effect insusceptible to intergalactic scattering dispersion, one of the limiting factors for FRB fringes~\cite{Zheng:2014rpa}. Compared with GRB fringes (also known as femtolensing, shown in \Fig{fig:results}), intensity correlation relaxes the source-size constraint, hence probing a much wider mass range. Lastly, the lensing parallax enables the detection of lensing with short pulses of GRBs. 

In all, the full range of PBH DM mass can be probed by lensing. GRB lensing parallax can uniquely probe the unconstrained lightest mass range, while star's lensing parallax can probe a heavier range complementary to existing observations. The maximum mass range is determined by detector separation, while the minimum depends crucially on GRB source size $\lesssim r_\odot$, high-energy spectrum  $\gtrsim$ 1-100 keV, and brightness resolution $\epsilon \lesssim 0.1$. It is also important to have a wide field of view such as in GBM in order to obtain high event rate. All these technologies are likely already available, and one (paired with existing ones) or two such space telescopes will be good enough to accomplish this probe. Fortunately, the astrophysical scale accessible to us -- from $r_\oplus$ to AU -- is just right to probe the lightest PBH mass range, which cannot be lensing probed otherwise.

\begin{acknowledgments}
Authors would like to thank Han Gil Choi, Myungshin Im, Ji-hoon Kim, Chang Dong Rho, Il H. Park, Sascha Trippe and Eunil Won for useful discussions. Authors are supported by Grant Korea NRF-2019R1C1C1010050, 2015R1A4A1042542, and SJ also by POSCO Science Fellowship.
\end{acknowledgments}

\appendix

\section{Useful numbers and formulas} 
Here we collect useful numbers and formulas, with $c=1$.

Cosmological constants we assumed are
\begin{subequations}
\bea
&&H_0 = 70 \text{km/s/Mpc}, \\
&&\rho_{crit, 0} = \frac{3 H_0^2}{8\pi G} = 9.21\times10^{-27} \ \text{kg/m}^3 \nonumber \\ &&\phantom{\rho_{crit, 0} = \frac{3 H_0^2}{8\pi G}} = 1.36\times10^{11} \Msun/\text{Mpc}^3, \\
&&\Omega_\Lambda = 0.7, \ \Omega_r = 0, \ \Omega_{M} = 0.3, \ \Omega_{DM} = 0.26.
\eea
\end{subequations}
Representative lengths are
\bea
r_\oplus \, &\simeq& \, 0.02 \, {\rm sec}, \quad r_\odot \,\simeq \, 2 \, {\rm sec}, \quad r_{L_2} \, \simeq \, 5 \, {\rm sec}, \nonumber\\
\quad {\rm AU} \, &\simeq&\, 500 \, {\rm sec}.
\eea
The Einstein radius of a PBH is
\beq
r_E \, \simeq \, 1.4 \times 10^6 \,{\rm sec} \, \sqrt{ \left( \frac{M}{\Msun} \right) \left( \frac{ D}{\rm Gpc} \right) }.
\eeq
The Schwarzschild radius is
\beq
r_{\rm Sch} \, \simeq \, 10^{-5} \, {\rm sec} \, \left( \frac{ M}{\Msun} \right).
\eeq
The time-delay is
\beq
\Delta t_d \, \sim \, 2 \times 10^{-5} \, {\rm sec} \, \left( \frac{M}{\Msun} \right).
\eeq

\section{Finite source-size effect} \label{app:sourcesize} 
We present the equation for the magnification $A$ that we used in the calculation, with finite source-size considered.

Magnification of a finite sized circular source with constant surface brightness by a point-lens in the geometrical optics limit is given by \cite{Witt:1994}
\beq
A (y, \delta) \approx 
\begin{cases}
A_{in}(y, \delta) & \text{for } y < \delta, \\
A_{out}(y, \delta) & \text{for } y > \delta,
\end{cases} \label{eq mag}
\eeq
where
\begin{subequations}
\begin{eqnarray}
A_{in}(y, \delta) &=& \sqrt{1+\frac{4}{\delta^2}}-\frac{8}{\delta^3 (\delta^2+4)^{3/2}}\frac{y^2}{2} \nonumber \\ 
&&-\frac{144(\delta^4+2\delta^2+2)}{\delta^5(\delta^2+4)^{7/2}}\frac{y^4}{24}, \\
A_{out}(y, \delta) &=& \frac{2+y^2}{y\sqrt{y^2+4}} +\frac{8(y^2+1)}{y^3(y^2+4)^{5/2}}\frac{\delta^2}{2} \nonumber \\
&&+\frac{48(3y^6+6y^4+14y^2+12)}{y^5(y^2+4)^{9/2}}\frac{\delta^4}{24},
\end{eqnarray}
\end{subequations}
and $\delta \equiv (r_S / D)/\theta_E$. Near $y \approx \delta $, $A_{in}$ and $A_{out}$ over- and underestimates the magnification compared to the actual value, respectively. In our calculation, we used each equation in $y\leq 0.9 \delta$ and $y \geq 1.1 \delta$ respectively, and then linearly interpolated the magnification in $0.9 \delta < y < 1.1 \delta$.

\section{Computing optical depth and constraints} 
Here we explain our computational procedure for the optical depths and the bounds on the PBH abundance. 

We assumed a uniform distribution of PBHs with number density $n$, so the optical depth for a source of physical transverse size $r_S$ at comoving distance $\chi_S$ is given by 
\bea
&&\tau(\chi_S; f; M; \Delta r_{\perp}, r_S, \epsilon) \nonumber \\ 
&&= n(f; M) \times V_L(\chi_S; \Delta r_{\perp}, r_S, \epsilon),
\eea
where $V_L$ is the desired comoving volume for PBHs and $\Delta r_{\perp} = \Delta r \cos \theta$ is the component of $\Delta r$ perpendicular to the line-of-sights (LOSs). The volume $V_L$ is given by 
\beq
V_L(\chi_S; \Delta r_{\perp}, r_S, \epsilon) = \int^{\chi_S}_0 \sigma(\chi_L; \Delta r_{\perp}, r_S, \epsilon) \ d\chi_L \label{eq:V_L}
\eeq
where $\sigma$ is the comoving cross-section for the lensing-detection of PBH at comoving distance $\chi_L$. Since the universe is flat, we do not distinguish the transverse comoving distance from the usual LOS comoving distance. 

To calculate $\sigma$, we first draw a region on the lens plane that at least one of the $A_1$ or $A_2$ exceeds $1+\epsilon$ (the necessary condition for Eq. (4) of the main paper). Then we draw a rectangular region circumscribing the previous one, and divide into 6400 equal-sized rectangular sectors (80 divisions for each edge). Next, we check whether Eq. (4) of the main paper is satisfied at the center of each sector, and summing up the area of the relevant sectors gives $\sigma$. 

Going back to \Eq{eq:V_L}, we equally divided the $\chi_L$ range into 24 sectors and used the midpoint quadrature for the integral. Finally, multiplying it by $n$ gives the optical depth $\tau$. Note that the only dependence on $f$ comes from $n \propto f$, so single evaluation of $\tau$ at $f=1$ immediately gives $\tau(f)$. 

Now we move onto the bound calculation. The probability of having no detection for observing $N$ sources is given by $P_{\rm null}(f) = \prod^N_{i=1} (1-P_{1, i}(f))$, where $i$ is the source index. The bound of $f$ is the one which gives $P_{\rm null}(f) = 0.05$. In the actual calculation, we sample 1000 sources, with distance and source size randomly selected according to the distributions given by \Fig{fig:GRBdistribution} and with purely random angular positions. We approximated $P_{\rm null}$ by assuming that each sampled source represents $N/1000$ observations, giving $P_{\rm null}(f) \approx \prod^{1000}_{i=1} (1-P_{1, i}(f))^{N/1000}$.

\section{Effects of PBH clustering} \label{app:clustering}
We now demonstrate that assuming a uniform distribution of PBHs is decent in calculating lensing parallax, even though nearly half of DM is thought to be clustered around galaxies. As discussed in the paper, we (conservatively) consider only single-lensing events; multi-lensing can still lead to parallax, but is more complicated to calculate and analyze. Thus, we show that the single lensing probability $P_1$ does not change significantly for clustered PBHs in most regions of galaxies.

First of all, the clustering does not change the optical depth $\tau$, as the average number of lenses within the $V_L$ (the desired volume of PBH locations for lensing) remains the same. But $P_1$ may still change because once a lens is within $V_L$ it is more likely that there are other clustered lenses within the same $V_L$ so that multi-lensing occurs more often than single lensing. More quantitatively, the number of lenses within $V_L$ no longer follows the Poisson distribution.

\begin{figure}[t] 
\centering
\includegraphics[width=0.48\textwidth]{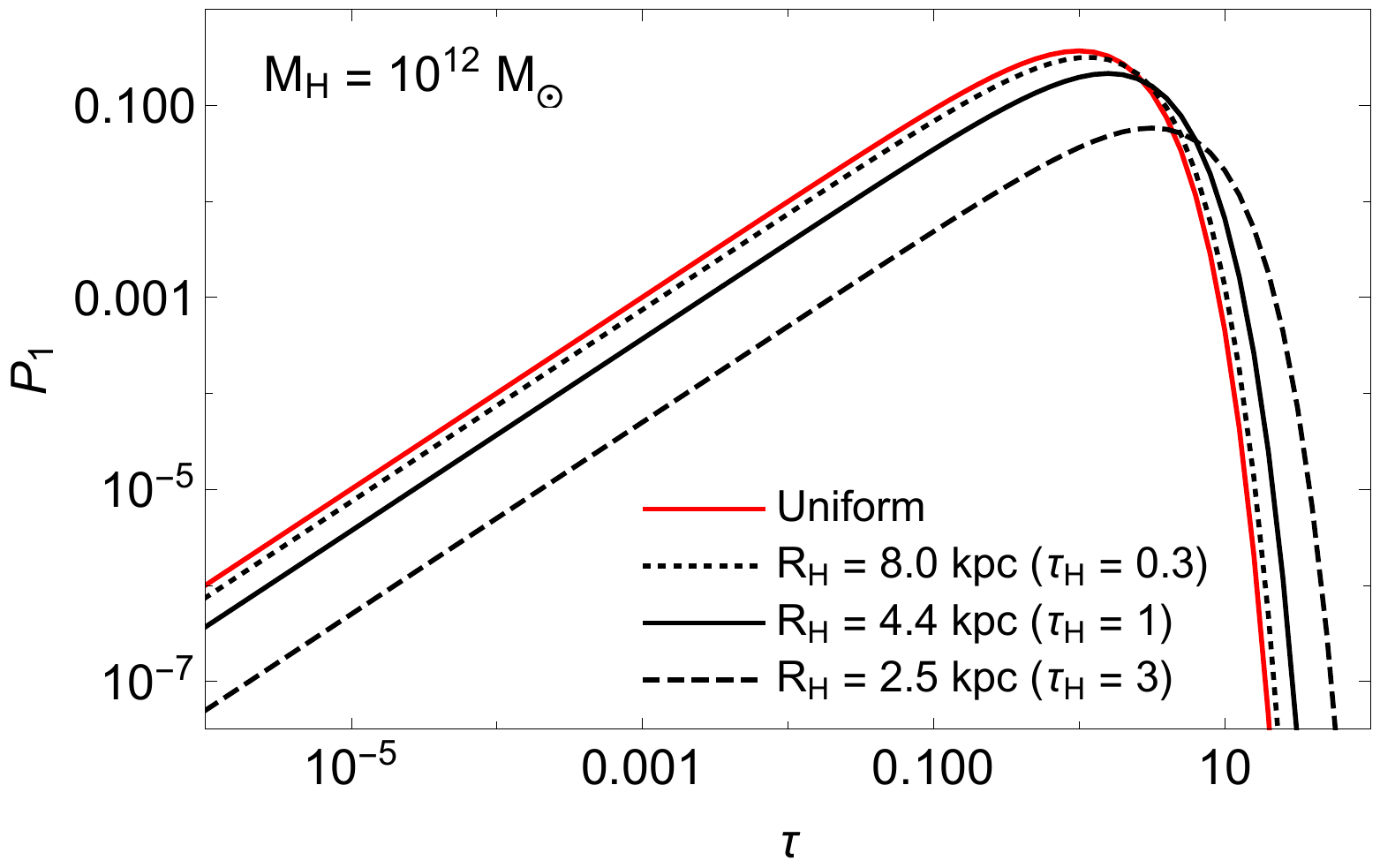}
\caption{Single lensing probability $P_1$ for selected values of $R_H$ (equivalently, $\tau_H$); the smaller the halo size $R_H$ of the given mass $M_H=10^{12}\Msun$, the higher clustering of PBHs within the single halo yielding higher optical depth $\tau_H$. The red line is for the uniform distribution of PBHs (no halos) giving the Poisson distribution of the number of lenses, while black lines represent clustered distributions which depart from the Poisson distribution. Note that the Milky Way has $R_H \sim 100$ kpc, yielding $\tau_H \ll 1$. $f=10^{-2}$ and $k=10$.}
\label{fig:clump}
\end{figure}

Suppose a LOS passes through $N$ halos (the clustered PBH) and the optical depth within each halo is denoted by $\tau_H$ (the more clustered within a halo, the larger $\tau_H$); then the expectation value of $\langle N \rangle = \tau / \tau_H$. The single-lensing probability for the case of $N$ halos is
\bea
&&P_{1|\text{N halos}} \nonumber \\ 
&&= (\text{number of halos}) \nonumber \\ 
&&\phantom{=} \times (\text{probability for one halo to give single lensing}) \nonumber \\ 
&&\phantom{=}\times (\text{probability for other halos to give no lensing}) \nonumber \\
&&= N\times \tau_H e^{-\tau_H} \times (e^{-\tau_H})^{N-1},
\eea
where we assume that PBH DM is clustered but uniformly distributed within each halo.
Summing $N$ with its own Poisson distribution, we obtain the total probability for single lensing
\beq
P_1 = \sum_{N=1}^{\infty} P_{1|\text{N halos}} \times \frac{(\tau/\tau_H)^N}{N!} e^{-\tau/\tau_H}. \label{eq:P1clump}
\eeq
\Fig{fig:clump} shows $P_1(\tau)$ for three selected values of $\tau_H$. The deviation from the uniform distribution is sizable for $\tau_H \gtrsim 1$ irrespective of $\tau$; this is the manifestation of non-Poissonian properties in \Eq{eq:P1clump}. This is understandable because $\tau_H \gtrsim 1$ directly means that clustering is so high that there are likely multiple lenses within the $V_L$ of a single halo.

The $\tau_H$ is related to the halo radius $R_H$ with the mass $M_H$.
The PBH number density within the halo is $n_{|H} = (M_H f / M)/(4/3 \times \pi R_H^3)$. The $V_L$ within the halo, denoted by $V_{L|H}$, is approximately the cylinder with a length $R_H$ and a (Einstein) cross-section $\sigma = k \cdot \pi (D_L \theta_E)^2$, where $k$ is determined by detection criteria and detector sensitivities (see below for realistic values of $k$). Thus we obtain $\tau_H$ as a function of $R_H$

\begin{equation}
\tau_H \approx 1.9 \times 10^{-22} \times k f  \left(\frac{M_H}{\Msun}\right)\left(\frac{D_L}{R_H}\right)^2\left(\frac{1 \ \text{Gpc}}{D_L D_S / D_{LS}}\right). \label{eq:tauH}
\end{equation}

This allows us to interpret \Fig{fig:clump} for Milky-Way-like galaxies. For $M_H = 10^{12}\, \Msun$, $\tau_H$ becomes $\gtrsim 1$ for $R_H \lesssim 4.4$ kpc. The Milky-Way is thought to have a much larger halo $\sim 100$ kpc. Thus, the PBH clustering is small enough not to affect our calculations based on the uniform distribution of PBH. For reference, the Milky-Way gives $\tau_H \lesssim 2 \times 10^{-3}$ with $R_H = 100$ kpc. In all these estimations, we use $k =10$ and $f= 10^{-2}$; the values of $k$ in our results are usually $\lesssim 10$ but grow with $\epsilon$ improvement so that the clustering and multi-lensing can become more relevant in the future.

\section{GRB source parameter distribution} \label{app:GRBparameter}
Finally, we present the distributions for GRB redshifts and transverse sizes that we use.

The GRB data on \cite{GRBdist} were used to obtain the redshift and the transverse size distributions for observed GRBs. Out of 2105 GRB observations, 487 have redshift information and 1944 have $T_{\rm 90}$ information which is converted to the transverse size by $r_S \sim c\times T_{\rm 90}$ \cite{Barnacka:2014yja}.

\begin{figure}[ht] 
\centering
\includegraphics[width=0.48\textwidth]{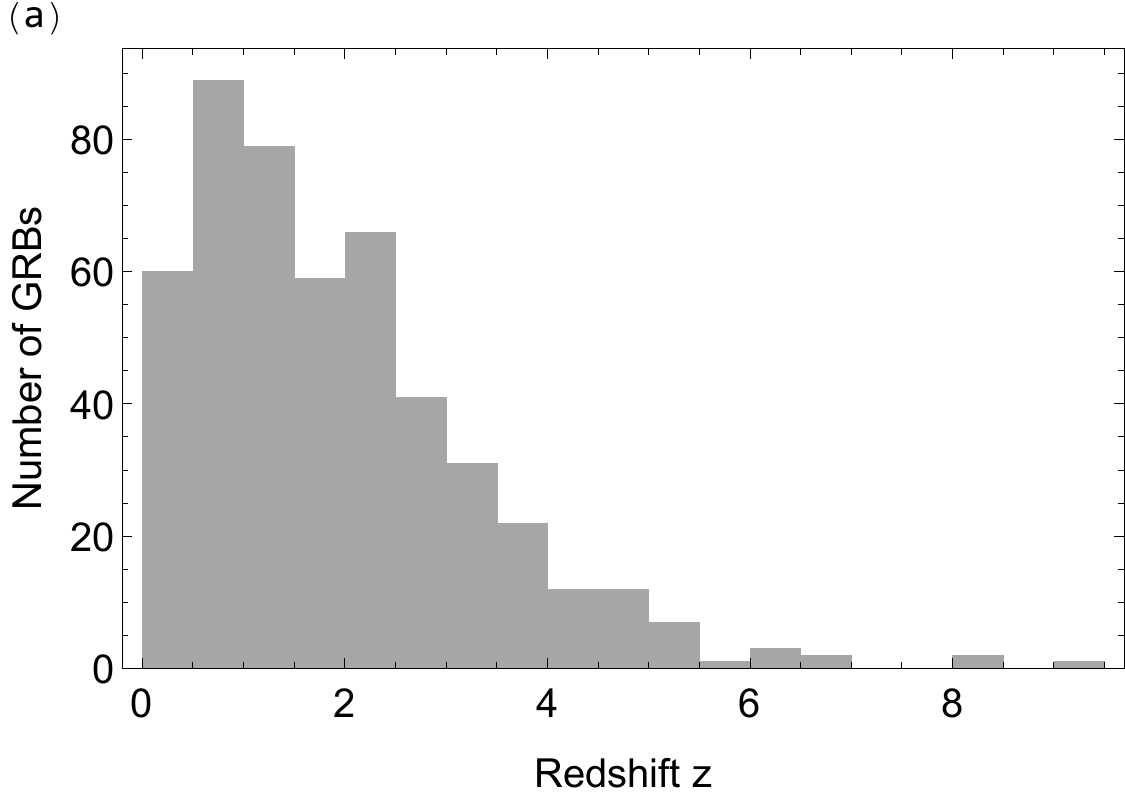}
\includegraphics[width=0.48\textwidth]{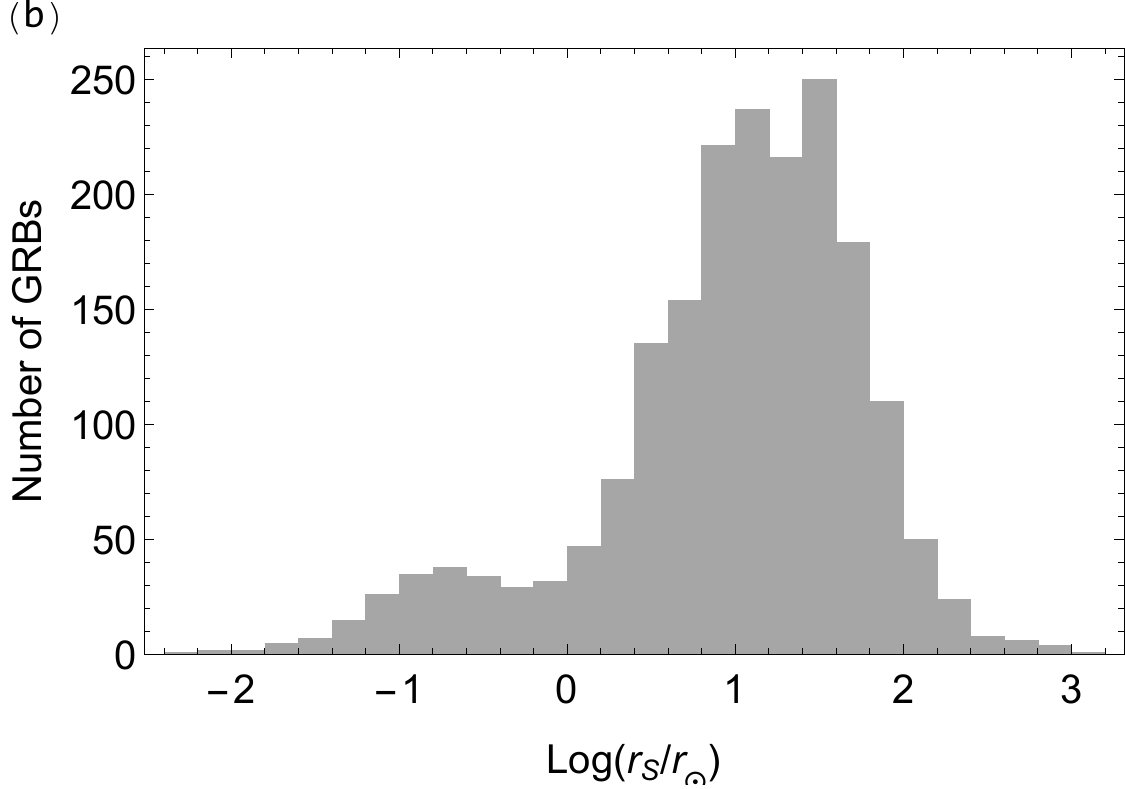}
\caption{Histograms of (a) redshift (upper panel) and (b) transverse size (lower panel) of observed GRBs that we use. Using a total of 2105 observed GRB data in \cite{GRBdist}, the redshift distribution was obtained from 487 GRBs with redshift information, and the transverse size distribution was obtained from 1944 GRBs with $T_{\rm 90}$ information that further converted by $r_S \sim c\times T_{\rm 90}$.
}
\label{fig:GRBdistribution}
\end{figure}

\Fig{fig:GRBdistribution} shows the redshift and transverse size distribution of observed GRBs. For the redshift distribution, the population extends to very high redshifts of $z\leq 10$, but the majority of the population lies in $z \lesssim 3$. For the transverse size distribution, the bimodal distribution is shown while the majority of the population lies in $r_S \geq \rsun$. Out of 1944 GRBs with transverse size converted from $T_{\rm 90}$, 226 (12\%) have $r_S \leq \rsun$ and only 58 (3\%) among them satisfies $r_S \leq 0.1 \rsun$. These fractions are smaller than the analysis done in \cite{Golkhou:2015lsa, Katz:2018zrn}, estimating about 10\% of observed GRBs have $r_S \leq 0.1\rsun$. Although the proper transverse size should be estimated by $T_{\rm 90}/(1+z_S)$, we omitted the redshift factor in calculating the size distribution to be conservative for unknown redshifts of the majority of GRBs used in obtaining the transverse size distribution.


\begin{thebibliography}{99}

\bibitem{Carr:2016drx} 
  B.~Carr, F.~Kuhnel and M.~Sandstad,
  ``Primordial Black Holes as Dark Matter,''
  Phys.\ Rev.\ D {\bf 94}, no. 8, 083504 (2016)
  doi:10.1103/PhysRevD.94.083504
  [arXiv:1607.06077 [astro-ph.CO]].

\bibitem{Sasaki:2018dmp} 
  M.~Sasaki, T.~Suyama, T.~Tanaka and S.~Yokoyama,
  ``Primordial black holes—perspectives in gravitational wave astronomy,''
  Class.\ Quant.\ Grav.\  {\bf 35}, no. 6, 063001 (2018)
  doi:10.1088/1361-6382/aaa7b4
  [arXiv:1801.05235 [astro-ph.CO]].
  
\bibitem{Niikura:2017zjd} 
  H.~Niikura {\it et al.},
  ``Microlensing constraints on primordial black holes with Subaru/HSC Andromeda observations,''
  Nat.\ Astron.\  {\bf 3}, no. 6, 524 (2019)
  doi:10.1038/s41550-019-0723-1
  [arXiv:1701.02151 [astro-ph.CO]].

\bibitem{Niikura:2019kqi} 
  H.~Niikura, M.~Takada, S.~Yokoyama, T.~Sumi and S.~Masaki,
  ``Constraints on Earth-mass primordial black holes from OGLE 5-year microlensing events,''
  Phys.\ Rev.\ D {\bf 99}, no. 8, 083503 (2019)
  doi:10.1103/PhysRevD.99.083503
  [arXiv:1901.07120 [astro-ph.CO]].

\bibitem{Griest:2013aaa} 
  K.~Griest, A.~M.~Cieplak and M.~J.~Lehner,
  Astrophys.\ J.\  {\bf 786}, no. 2, 158 (2014)
  doi:10.1088/0004-637X/786/2/158
  [arXiv:1307.5798 [astro-ph.CO]].

\bibitem{Tisserand:2006zx} 
  P.~Tisserand {\it et al.} [EROS-2 Collaboration],
  Astron.\ Astrophys.\  {\bf 469}, 387 (2007)
  doi:10.1051/0004-6361:20066017
  [astro-ph/0607207].

\bibitem{Allsman:2000kg} 
  R.~A.~Allsman {\it et al.} [Macho Collaboration],
  Astrophys.\ J.\  {\bf 550}, L169 (2001)
  doi:10.1086/319636
  [astro-ph/0011506].

\bibitem{Dror:2019twh} 
  J.~A.~Dror, H.~Ramani, T.~Trickle and K.~M.~Zurek,
  ``Pulsar Timing Probes of Primordial Black Holes and Subhalos,''
  Phys.\ Rev.\ D {\bf 100}, no. 2, 023003 (2019)
  doi:10.1103/PhysRevD.100.023003
  [arXiv:1901.04490 [astro-ph.CO]].

\bibitem{Zumalacarregui:2017qqd} 
  M.~Zumalacarregui and U.~Seljak,
  ``Limits on stellar-mass compact objects as dark matter from gravitational lensing of type Ia supernovae,''
  Phys.\ Rev.\ Lett.\  {\bf 121}, no. 14, 141101 (2018)
  doi:10.1103/PhysRevLett.121.141101
  [arXiv:1712.02240 [astro-ph.CO]].

\bibitem{Munoz:2016tmg} 
  J.~B.~Muñoz, E.~D.~Kovetz, L.~Dai and M.~Kamionkowski,
  ``Lensing of Fast Radio Bursts as a Probe of Compact Dark Matter,''
  Phys.\ Rev.\ Lett.\  {\bf 117}, no. 9, 091301 (2016)
  doi:10.1103/PhysRevLett.117.091301
  [arXiv:1605.00008 [astro-ph.CO]].

\bibitem{Jung:2017flg} 
  S.~Jung and C.~S.~Shin,
  ``Gravitational-Wave Fringes at LIGO: Detecting Compact Dark Matter by Gravitational Lensing,''
  Phys.\ Rev.\ Lett.\  {\bf 122}, no. 4, 041103 (2019)
  doi:10.1103/PhysRevLett.122.041103
  [arXiv:1712.01396 [astro-ph.CO]].
  
\bibitem{Katz:2018zrn} 
  A.~Katz, J.~Kopp, S.~Sibiryakov and W.~Xue,
  ``Femtolensing by Dark Matter Revisited,''
  JCAP {\bf 1812}, 005 (2018)
  doi:10.1088/1475-7516/2018/12/005
  [arXiv:1807.11495 [astro-ph.CO]].

\bibitem{Bai:2018bej} 
  Y.~Bai and N.~Orlofsky,
  ``Microlensing of X-ray Pulsars: a Method to Detect Primordial Black Hole Dark Matter,''
  Phys.\ Rev.\ D {\bf 99}, no. 12, 123019 (2019)
  doi:10.1103/PhysRevD.99.123019
  [arXiv:1812.01427 [astro-ph.HE]].
  
\bibitem{Nemiroff:1995ak} 
  R.~J.~Nemiroff and A.~Gould,
  ``Probing for MACHOs of mass $10^{-15} \Msun$--$10^{-7} \Msun$ with gamma-ray burst parallax spacecraft,''
  Astrophys.\ J.\  {\bf 452}, L111 (1995)
  doi:10.1086/309722
  [astro-ph/9505019].

\bibitem{Gould:1992}
A.~Gould,
``Femtolensing of gamma-ray bursters,''
  Astrophys.\ J.\  {\bf 386}, L5 (1992)
  doi:10.1086/186279
  
\bibitem{Nakamura:1997sw} 
  T.~T.~Nakamura,
  ``Gravitational lensing of gravitational waves from inspiraling binaries by a point mass lens,''
  Phys.\ Rev.\ Lett.\  {\bf 80}, 1138 (1998).
  doi:10.1103/PhysRevLett.80.1138
  
\bibitem{Witt:1994} 
  H.~J.~Witt and S.~Mao,
  ``Can lensed stars be regarded as pointlike for microlensing by MACHOs?,''
  Astrophys.\ J.\  {\bf 430}, no. 2, 505 (1994)
  doi:10.1086/174426
  
\bibitem{Takahashi:2003ix} 
  R.~Takahashi and T.~Nakamura,
  ``Wave effects in gravitational lensing of gravitational waves from chirping binaries,''
  Astrophys.\ J.\  {\bf 595}, 1039 (2003)
  doi:10.1086/377430
  [astro-ph/0305055].
  
\bibitem{Barnacka:2014yja} 
  A.~Barnacka and A.~Loeb,
  ``A size-duration trend for gamma-ray burst progenitors,''
  Astrophys.\ J.\  {\bf 794}, no. 1, L8 (2014)
  doi:10.1088/2041-8205/794/1/L8
  [arXiv:1409.1232 [astro-ph.HE]].

\bibitem{GRBdist}
\url{http://www.astro. caltech.edu/grbox/grbox.php}

\bibitem{Golkhou:2015lsa} 
  V.~Z.~Golkhou, N.~R.~Butler and O.~M.~Littlejohns,
  ``The Energy-Dependence of GRB Minimum Variability Timescales,''
  Astrophys.\ J.\  {\bf 811}, no. 2, 93 (2015)
  doi:10.1088/0004-637X/811/2/93
  [arXiv:1501.05948 [astro-ph.HE]].

\bibitem{TypicalSpectrum}
\url{https://fermi.gsfc.nasa.gov/ssc/data/analysis/documentation/Cicerone/Cicerone_GRBs/Overview_GRB_Spec_Anal.html}

The shown typical spectrum peaks at ${\cal O}(100)$ keV with energy flux $\sim 1$ MeV cm$^{-2}$ sec$^{-1}$; a typo in the unit is corrected here. With 100 cm$^2$ Fermi detector area, ${\cal O}(100)$ photon counts per MeV is typically achieved at ${\cal O}(100)$ keV.


\bibitem{Carr:2009jm} 
  B.~J.~Carr, K.~Kohri, Y.~Sendouda and J.~Yokoyama,
  Phys.\ Rev.\ D {\bf 81}, 104019 (2010)
  doi:10.1103/PhysRevD.81.104019
  [arXiv:0912.5297 [astro-ph.CO]].

\bibitem{Ali-Haimoud:2016mbv} 
  Y.~Ali-Haïmoud and M.~Kamionkowski,
  Phys.\ Rev.\ D {\bf 95}, no. 4, 043534 (2017)
  doi:10.1103/PhysRevD.95.043534
  [arXiv:1612.05644 [astro-ph.CO]].
  
\bibitem{Refsdal:1986} 
  B.~Grieger, R.~Kayser, S.~Refsdal,
  ``A parallax effect due to gravitational micro-lensing,''
  Nature\  {\bf 324}, 126 (1986)
  doi:10.1038/324126a0.
  
\bibitem{Bachelet:2018} 
  E.~Bachelet, T.~C.~Hinse, R.~Street,
  ``Measuring the Microlensing Parallax from Various Space Observatories,''
  Astrophys.\ J.\  {\bf 155}, no. 5, 7 (2018)
  doi:10.3847/1538-3881/aab3c8
  [arXiv:1803.00689 [astro-ph.EP]].
  
\bibitem{Zheng:2014rpa} 
  Z.~Zheng, E.~O.~Ofek, S.~R.~Kulkarni, J.~D.~Neill and M.~Juric,
  Astrophys.\ J.\  {\bf 797}, no. 1, 71 (2014)
  doi:10.1088/0004-637X/797/1/71
  [arXiv:1409.3244 [astro-ph.HE]].



\end{thebibliography}
\end{document}